\documentclass[review]{elsarticle}

\usepackage{hyperref}
\usepackage{amsmath}
\usepackage{multirow}
\usepackage{placeins}

\journal{Physica A: Statistical Mechanics and its Applications}

\usepackage{xr}
\begin{document}

\begin{frontmatter}

\title{Temporal stability in human interaction networks\tnoteref{mytitlenote}}
\tnotetext[mytitlenote]{The Supporting Information document supplies thorough tables and figures.}

\author[add1]{Renato Fabbri\corref{mycorrespondingauthor}}
\address[add1]{S\~ao Carlos Institute of Physics, University of S\~ao Paulo (IFSC/USP)}
\ead{fabbri@usp.br}

\author[add2]{Ricardo Fabbri}%
\address[add2]{Polytechnic Institute, Rio de Janeiro State University (IPRJ/UERJ)}
\ead{rfabbri@iprj.uerj.br}

\author[add3]{Deborah Christina Antunes}
\ead{deborahantunes@gmail.com}
\address[add3]{Psychology Course, Federal University of Cear\'a (UFC), Sobral Campus}

\author[add4]{Marilia Mello Pisani}
\ead{marilia.m.pisani@gmail.com}
\address[add4]{Center for Natural and Human Sciences, Federal University of ABC (CCNH/UFABC)}

\author[add1]{Osvaldo Novais de Oliveira Junior}
\ead{chu@ifsc.usp.br}

\cortext[mycorrespondingauthor]{Corresponding author}

\begin{abstract}
This paper reports on stable (or invariant) properties of human interaction networks,
with benchmarks derived from public email lists. Activity, recognized through messages sent,
along time and topology were observed in snapshots in a timeline, 
and at different scales. Our analysis shows that activity is practically the same for all networks across timescales ranging from seconds to months.
The principal components of the participants in the topological metrics space remain practically unchanged as different sets of messages are considered.
The activity of participants follows the expected scale-free trace, thus yielding the hub, intermediary and peripheral classes of vertices by comparison against the Erd\"os-R\'enyi model. The relative sizes of these three sectors are essentially the same for all email lists and the same along time.
Typically, $<15\%$ of the vertices are hubs, 15-45\% are intermediary and $>45\%$ are peripheral vertices. Similar results for the distribution of participants in the three sectors and for the relative importance of the topological metrics were obtained for 12 additional networks from Facebook, Twitter and ParticipaBR. These properties are consistent with the literature and may be general for human interaction networks, which has important implications for establishing a typology of participants based on quantitative criteria.
\end{abstract}


\end{frontmatter}

\begin{quotation}
`The reason for the persistent plausibility of the typological approach, however, is not a static biological one, but just the opposite: dynamic and social.' 
\emph{- Adorno et al, 1969, p. 747}
\end{quotation}

\section{Introduction}\label{sec:into}
The first studies dealing explicitly with human interaction networks
date from the nineteenth century while the foundation of
social network analysis is generally attributed to the psychiatrist Jacob Moreno in mid twentieth century~\cite{moreno,newmanBook}. With the increasing availability of data related to human interactions, research about these networks has grown continuously. Contributions can now be found in a variety of fields, from social sciences and humanities~\cite{latour2013} to computer science~\cite{bird} and physics~\cite{barabasiHumanDyn,newmanFriendship}, given the multidisciplinary nature of the topic. One of the approaches from an exact science perspective is to represent interaction networks as complex networks~\cite{barabasiHumanDyn,newmanFriendship}, with which 
several features of human interaction have been revealed. For example, the topology of human interaction networks exhibits a scale-free trace, which points to the existence of a small number of highly connected hubs and a large number of poorly connected nodes. The dynamics of complex networks representing human interaction has also been addressed~\cite{barabasiEvo,newmanEvolving}, but only to a limited extent, since research is normally focused on a particular metric or task, such as accessibility or community detection~\cite{access,newmanModularity}. 

In this paper we analyze the evolution of human interaction networks.
Directed and weighted representations were built through the observation of replies as links.
Interaction networks from email lists were the most convenient for deriving results and for benchmarking while networks from Facebook, Twitter and ParticipaBR were used for the sake of generalization.
Using a timeline of activity snapshots with a constant number of contiguous messages, we found remarkable stability (or invariance) for important network properties. For instance, activity along different timescales follows specific patterns; the most basic topological metrics can always be combined into characteristic principal components; and the fractions of participants in different sectors do not vary with time. This is not an intuitive result, given that participants constantly transition in network structure. Because these properties were shared by networks from various sources, and are consistent with the literature in complex networks~\cite{newmanBook}, we advocate that the conclusions might be valid for general classes of interaction networks. In particular, this allows us to 
bridge the gap between data analysis and social sciences in the discussion of types of networks and of participants.
It is worth noting that typologies are the canon of scientific literature for the classification of human agents, with pragmatic standards~\cite{myers} and critical paradigms~\cite{adorno,typCanon}. 

This paper is organized as follows. Section~\ref{sec:related} describes related work, while data, scripts and methods of analysis are given in Section~\ref{sec:data} and Section~\ref{sec:carac}.
Section~\ref{sec:results} reports results and discussion, leading to Section~\ref{sec:conc} for conclusions.
Supplementary data analysis, including directions for video and sound mappings of network structures, and numeric detailed results for networks from Twitter, Facebook and ParticipaBR, are provided in the Supporting Information document.

\subsection{Related work}\label{sec:related}
The fact that unreciprocated edges often exceed 50\% in human interaction networks~\cite{newmanEvolving} motivated the inclusion of symmetry metrics in our analysis.
No correlation of topological characteristics and geographical coordinates was found~\cite{barabasiGeo},
therefore geographical positions were not considered in our study.
Gender related behavior in mobile phone datasets was indeed reported~\cite{barabasiSex}
but it is not relevant for the present work because email messages and addresses have no gender related metadata~\cite{gmanePack}.

Research on network evolution is often restricted to network growth, in which there is a monotonic increase in the number of events~\cite{barabasiEvo}.
Network types have been discussed with regard to the number of participants, intermittence of their activity and network longevity~\cite{barabasiEvo}. Two topologically different networks emerged from human interaction networks, depending on whether the frequency of interactions follows a generalized power law or an exponential connectivity distribution~\cite{barabasiTopologicalEv}. In email list networks, scale-free properties were reported with $\alpha \approx 1.8$~\cite{bird} (as in web browsing and library loans~\cite{barabasiHumanDyn}), and different linguistic traces were related to weak and strong ties~\cite{Gmane2}.

\section{Data and scripts}\label{sec:data}\label{scripts}

Email list messages were obtained from
the Gmane email archive, which consists of more than $20,000$
email lists (discussion groups) and more than $130\times 10^6$ messages~\cite{Gmanewikipedia}. These lists cover a variety of topics, mostly technology-related. The archive can be described as a corpus along with message metadata, including sent time, place, sender name, and sender email address.
The usage of the Gmane database in scientific research is reported in studies of isolated lists and of lexical innovations~\cite{Gmane2,bird}. 

We observed various email lists and selected four of them together with data from Twitter, Facebook and ParticipaBR for a thorough analysis,
from which general properties can be inferred. These lists are as follows:

\begin{itemize}
\item Linux Audio Users list\footnote{gmane.linux.audio.users is list ID in Gmane.},with participants from different countries with artistic and technological interests. English is the prevailing language. Abbreviated as LAU from now on.

\item Linux Audio Developers list\footnote{gmane.linux.audio.devel is list ID in Gmane.}, with participants from different countries; a more technical and less active version of LAU. English is the prevailing language. Abbreviated as LAD from now on.

\item Developer's list for the standard C++ library\footnote{gmane.comp.gcc.libstdc++.devel is list ID in Gmane.}, with computer programmers from different countries. English is the prevailing language. Abbreviated as CPP from now on.
\item List of the MetaReciclagem project\footnote{gmane.politics.organizations.metareciclagem is list ID in Gmane.}, a Brazilian email list for digital culture. 	Portuguese is the prevailing language, although some messages are written in Spanish and English. Abbreviated as MET from now on.
\end{itemize} 

The first 20,000 messages of each list were considered, with basic attributes of total timespan, authors, threads and missing messages indicated in Table~\ref{tab:genLists}. We considered 140 additional email lists to report on the interdependence between the number of participants and the number of discussion threads. Furthermore, 12 networks from Facebook (8), Twitter (2) and ParticipaBR (2) were scrutinized, and their analysis is given in the Supporting Information document for the purpose of testing the generality of the results.

\begin{table}
\centering
\label{tab:genLists}
\begin{tabular}{l|c|c|c|c|c}\hline
list & $date_1$ & $date_{M}$    & $N$  & $\Gamma$ & $\overline{M}$ \\\hline
LAU & 2003-06-29  & 2005-07-23  & 1147  & 3374  & 5 \\
LAD & 2003-07-03  & 2009-10-07  & 1232  & 3114  & 4 \\
MET & 2005-08-01  & 2008-03-07  & 477  & 4607  & 23 \\
CPP & 2002-03-12  & 2009-08-25  & 1036  & 4506  & 7 \\\hline

\end{tabular}
	\caption{{\bf Overview of the email lists analyzed.} Columns $date_1$ and $date_M$ have dates of first and last messages from the 20,000 messages considered in each email list.
$N$ is the number of participants (number of different email addresses),
$\Gamma$ is the number of discussion threads (count of messages without antecedent),
$\overline{M}$ is the number of messages missing in the 20,000 collection
($100\frac{23}{20000}=0.115$ percent in the worst case).
}
\end{table}

The data and scripts used to derive the results, figures and tables, and this article itself are publicly available. Email messages are downloadable from the Gmane public database~\cite{Gmanewikipedia}.
Data annotated from Facebook and Twitter are in a public repository~\cite{fbtwData}.
Data from ParticipaBR were used from the linked data/semantic web RDF triples~\cite{opa}, available in~\cite{datahub}.
Computer scripts are delivered through a public domain Python PyPI package and an open Git repository~\cite{gmanePack}.
This open approach to both data and scripts reinforces the scientific aspect of the contribution~\cite{openSci} and mitigates ethical and moral issues involved in researching systems constituted of human individuals~\cite{anPhy,ccs15}.

\section{Methods}\label{sec:carac}
%The networks were characterized with: 1) statistics of activity along time, in scales from seconds to years; 2) dispersion of basic topological metrics; 3) sectioning of the networks in hubs, intermediary and periphery; 4) iterative visualization, sonification and data inspection.
%These procedures are described below.

\subsection{Temporal activity statistics}\label{sec:mtime}
Messages were counted over time as histograms in the scales of seconds,
minutes, hours, days of the week, days of the month, and months of the year.
Most standard measures of location and dispersion, e.g. the usual mean and
standard deviation, hold little meaning in a compact Riemannian manifold,
such as the recurrent time periods that we are interested in.
Similar measures were taken using circular statistics~\cite{directionalStats},
in which each measurement $t$ is represented as a unit complex number,
$z=e^{i\theta}=\cos(\theta)+i\sin(\theta)$, where $\theta=t\frac{2\pi}{T}$,
and $T$ is the period in which the counting is repeated.
For example, $\theta=12\frac{2\pi}{24}=\pi$ for a message sent at $t=12h$ and given $T=24h$ for days.
The moments $m_n$, lengths of moments $R_n$, mean angles $\theta_\mu$, and rescaled mean angles $\theta_\mu'$ are defined as:

\begin{align}\label{eq:cmom}
m_n&=\frac{1}{N}\sum_{i=1}^N z_i^n \nonumber\\
R_n&=|m_n|\\
\theta_\mu&=Arg(m_1) \nonumber \\
\theta_\mu'&=\frac{T}{2\pi} \theta_\mu \nonumber
\end{align}

$\theta_\mu'$ is used as the measure of location.
Dispersion is measured using the circular variance $Var(z)$,
the circular standard deviation $S(z)$, and the circular dispersion $\delta(z)$:

\begin{align}\label{eq:cmd}
Var(z)&=1 - R_1 \nonumber\\
S(z)&= \sqrt{-2\ln(R_1)}\\
\delta(z)&=\frac{1-R_2}{2 R_1^2} \nonumber
\end{align}

\noindent
Also, the ratio $r=\frac{b_l}{b_h}$ between the lowest $b_l$ and the highest $b_h$ incidences on the histograms 
served as a further clue of how close the distribution was to being uniform. As expected, a positive correlation was found in all $r, Var(z)$, $S(z)$ and $\delta(z)$ dispersion measures,
which can be noticed in Section~\ref*{si:circ} of the Supporting Information. The circular dispersion $\delta(z)$ was found more sensitive and therefore preferred in the discussion of results.

\subsection{Interaction networks}\label{intNet}
Edges in interaction networks can be modeled both as weighted or unweighted, as directed or undirected~\cite{bird,newmanCommunityDirected,newmanCommunity2013}.
Networks in this paper are directed and weighted, the most informative of the possibilities. We did not investigate directed unweighted, undirected weighted, and undirected unweighted representations of the interaction networks. 

\begin{figure}[!h]
\centering
\includegraphics[width=0.4\textwidth]{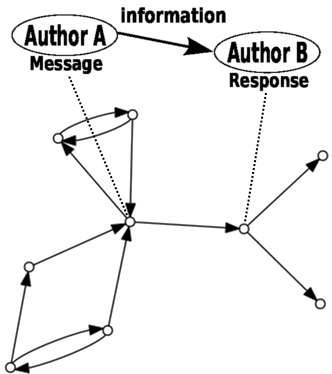}
	\caption{{\bf The formation of interaction networks from exchanged messages.} Each vertex represents a participant. A reply message from author B to a message from author A is regarded as evidence that B received information from A and yields a directed edge. 	Multiple messages add ``weight'' to a directed edge. Further details are given in Section~\ref{intNet}.}
\label{formationNetwork}
\end{figure}

The interaction networks were obtained as follows: a direct response from participant B to a message from participant A yields an edge from A to B, as information went from A to B. The reasoning is: if B wrote a response to a message from A, he/she read what A wrote and formulated a response, so B assimilated information from A, thus $A \rightarrow B$.
Edges in both directions are allowed. Each time an interaction occurs, the value of one is added to the edge weight. Selfloops were regarded as non-informative and discarded. Inverting edge direction yields the status network: B read the message and considered what A wrote worth responding, giving status to A, thus $B\rightarrow A$. This paper considers by convention the information network as described above ($A\rightarrow B$) and depicted in Figure~\ref{formationNetwork}. These interaction networks are reported in the literature as exhibiting scale-free and small-world properties, as expected for a number of social networks~\cite{bird,newmanBook}.

%Edges can be created from all antecedent message authors on the message-response thread to each message author.
%We only linked the immediate antecedent to the new message author, both for simplicity and because in adding two edges, $x\rightarrow y$ and $y\rightarrow z$, there is also a weaker connection between $x$ and $z$. Potential interpretations for this weaker connection are: double length, half weight or with one more ``obstacles''. This suggests the adequacy of centrality measurements to account for the connectivity of a node with all other nodes, such as betweenness centrality, eigenvector centrality and accessibility~\cite{luMeasures,access}.

\subsubsection{Topological metrics}\label{measures}

The topology of the networks was characterized 
from a small selection of the most basic 
and fundamental measurements for each vertex~\cite{newmanBook}, as follows:

\begin{itemize}
\item Degree     $k_i$: number of edges linked to vertex $i$.
\item In-degree  $k_i^{in}$: number of edges ending at vertex $i$.
\item Out-degree $k_i^{out}$: number of edges departing from vertex $i$.
\item Strength $s_i$: sum of weights of all edges linked to vertex $i$.
\item In-strength $s_i^{in}$: sum of weights of all edges ending at vertex $i$.
\item Out-strength $s_i^{out}$: sum of weights of all edges departing from vertex $i$.
\item Clustering coefficient $cc_i$: fraction of pairs of neighbors of $i$ that are linked, i.e. the standard clustering coefficient metric for undirected graphs.
\item Betweenness centrality $bt_i$: fraction of geodesics that contain vertex $i$. The betweenness centrality index was computed for weighted digraphs as specified in~\cite{faster}.
\end{itemize}

The non-standard metrics below were formulated to capture symmetries in the activity of participants:

\begin{itemize}
\item Asymmetry of vertex $i$: $asy_i=\frac{k_i^{in}-k_i^{out}}{k_i}$.
\item Average asymmetry of edges at vertex $i$:\\ $\mu_i^{asy}=\frac{\sum_{j\in J_i} e_{ji}-e_{ij}}{|J_i|}$, where $e_{ij}$ is 1 if there is an edge from $i$ to $j$, and $0$ otherwise, and $J_i$ is the set of neighbors of vertex $i$.
\item Standard deviation of asymmetry of edges:\\ $\sigma_i^{asy}=\sqrt{\frac{\sum_{j\in J_i}[\mu^{asy}_i -(e_{ji}-e_{ij}) ]^2  }{|J_i|}  }$.
\item Disequilibrium: $dis_i=\frac{s_i^{in}-s_i^{out}}{s_i}$.
\item Average disequilibrium of edges:\\ $\mu_i^{dis}=\frac{\sum_{j \in J_i}\frac{w_{ji}-w_{ij}}{w_{ji}+w_{ij}}}{|J_i|}$, where $w_{xy}$ is the weight of edge $x\rightarrow y$ and zero if there is no such edge.
\item Standard deviation of disequilibrium of edges: $\sigma_i^{dis}=\sqrt{\frac{\sum_{j\in J_i}\left[\mu^{dis}_i-\frac{w_{ji}-w_{ij}}{w_{ji}+w_{ij}}\right]^2}{|J_i|}}$.
\end{itemize}

Both standard and non-standard metrics are used for the Erd\"os sectioning (described in Section~\ref{sectioning}) and for performing principal component analysis (PCA) (as described in Section~\ref{sec:pca}).

\subsection{Erd\"os sectioning}\label{sectioning}
\begin{figure}[!h]
\centering
\includegraphics[width=\textwidth]{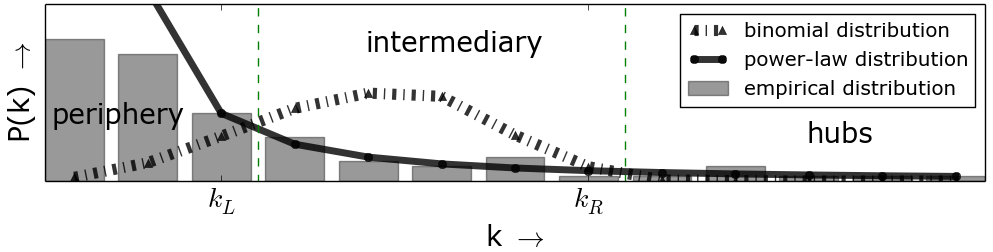}
	\caption{{\bf Three sectors of a scale-free networks.} This is a classification of vertices by comparing degree distributions~\cite{3setores}.
The binomial distribution of the Erd\"os-R\'enyi network model exhibits more intermediary vertices, while a scale-free network, associated with the power-law distribution, has more peripheral and hub vertices. The sector borders are defined with respect to the intersections of the distributions. Characteristic degrees are in the compact intervals: $[0,k_L]$, $(k_L,k_R]$, $(k_R,k_{max}]$ for the periphery, intermediary and hub sectors, the ``Erd\"os sectors''.
The connectivity distribution of empirical interaction networks, e.g. derived from email lists, can be sectioned by comparison against the associated binomial distribution with the same number of vertices and edges. In this figure, a snapshot of 1000 messages from CPP list yields the degree distribution of an interaction network of 98 nodes and 235 edges. A thorough explanation of the method is provided in Section~\ref{sectioning}.}
\label{fig:setores}
\end{figure}

It is often useful to think of vertices as hubs, peripheral and intermediary. We have therefore derived the peripheral, intermediary and hub sectors of the empirical networks from a comparison against an Erd\"os-R\'enyi network with the same number of edges and vertices,
as depicted in Figure~\ref{fig:setores}. We refer to this procedure as \emph{Erd\"os sectioning}, with the resulting sectors being named as \emph{Erd\"os sectors}. The Erd\"os sectioning was recognized as a theoretical possibility by M. O. Jackson in his video lectures~\cite{3setores}, but to our knowledge it has not as yet been applied to empirical data.

The degree distribution $\widetilde{P}(k)$ of a real network with a scale-free profile $\mathcal{N}_f(N,z)$ with $N$ vertices and $z$ edges has less
average degree nodes than the distribution $P(k)$ of an Erd\"os-R\'enyi
network with the same number of vertices and edges. Indeed, we define in this work the intermediary sector of a network to be the set of all the nodes whose degree is less abundant in the real network than on the Erd\"os-R\'enyi model:

\begin{equation}\label{criterio}
\widetilde{P}(k)<P(k) \Rightarrow \text{k is intermediary degree}
\end{equation}

If $\mathcal{N}_f(N,z)$ is directed and has no self-loops, the probability of the existence
of an edge between two arbitrary vertices is $p_e=\frac{z}{N(N-1)}$.
A vertex in the ideal Erd\"os-R\'enyi digraph with the same number of vertices and edges, and thus the same probability $p_e$ for the presence of an edge, will have degree $k$ with probability

\begin{equation}
P(k)=\binom{2(N-1)}{k}p_e^k(1-p_e)^{2(N-1)-k}
\end{equation}

The lower degree fat tail corresponds to the border vertices, i.e. the peripheral sector or periphery where $\widetilde{P}(k)>P(k)$ and $k$ is lower than any value of $k$ in the intermediary sector.
The higher degree fat tail is the hub sector, i.e. $\widetilde{P}(k)>P(k)$ and $k$ is higher than any value of $k$ in the intermediary sector. The reasoning for this classification is as follows: vertices so connected that they are virtually nonexistent in the Erd\"os-R\'enyi model, are coherently associated to the hub sector.
Vertices with very few connections, which are way more abundant than expected in the Erd\"os-R\'enyi model,
are assigned to the periphery.
Vertices with degree values predicted as the most abundant in the Erd\"os-R\'enyi model,
near the average, and less frequent in the real network, are classified as intermediary.

\clearpage

To ensure statistical validity of the histograms, bins can be chosen to contain at least $\eta$ vertices of the real network.
The range $\Delta$ of incident values of degree $k$ should be partitioned in $m$ parts $\Delta=\cup_{i=1}^m \Delta_i$,
with $\Delta_i\cap \Delta_j=\emptyset \; \forall\; i \neq j$ and:
%\begin{equation}
%\begin{split}
\begin{align}
\Delta_i =\Biggl\{ k\;\; & | & \overline{\Delta}_{i-1}< &\, k\leq l \text{ and }\;\;\;\;\;\;\;\;\;\;\;\;\nonumber\\
                         & \Biggl[ \Bigl[ & N - \sum_{k=0}^{\overline{\Delta}_{i-1}} & \eta_k< \eta \text{ and } l = \overline{\Delta} \Bigr] \text{ or }\\
&	\Bigl[ & \sum_{k=\overline{\Delta}_{i-1}+1}^l &  \eta_k \geq \eta \text{ and }\;\;\;\;\;\;\nonumber\\
& \Bigl( &\sum_{k=\overline{\Delta}_{i-1}+1}^{l-1} &  \eta_k < \eta \text{ or }   l=\overline{\Delta}_{i-1}+1 \Bigr) \;\Bigr] \Biggr] \Biggr\}\nonumber
\end{align}
%\end{split}
%\end{equation}

\noindent where $\eta_k$ is the number of vertices with degree $k$,
while $\overline{\Delta}_{(i)}=max(\Delta_{(i)})$, and $\overline{\Delta}_{0}=-1$.
% and $\overline{\Delta}_{i}<l\leq max(\Delta)$.
Equation~\ref{criterio} can now be written in the form:

\begin{equation}\label{criterio2}
\begin{split}
\sum_{x=min(\Delta_i)}^{\overline{\Delta}_i} \widetilde{P}(x) < \sum_{x=min(\Delta_i)}^{\overline{\Delta}_i} P(x) \Leftrightarrow \\
\Leftrightarrow \Delta_i \text{ spans intermediary degree values.}
\end{split}
\end{equation}

If the strength $s$ is used for comparison of the real network against the Erd\"os-R\'enyi model,
$P$ remains the same, but $P(\kappa_i)$ with $\kappa_i=\frac{s_i}{\overline{w}}$ should be used, where $\overline{w}=2\frac{z}{\sum_is_i}$ is the average weight of an edge and $s_i$ is the strength of vertex $i$. For in and out degrees ($k^{in}$, $k^{out}$), the real network should be compared against
\begin{equation}
\hat{P}(k^{way})=\binom{N-1}{k^{way}}p_e^k(1-p_e)^{N-1-k^{way}},
\end{equation}

\noindent where \emph{way} can be \emph{in} or \emph{out}. In and out strengths ($s^{in}$, $s^{out}$) are divided by $\overline{w}$ and compared also using $\hat{P}$. Note that $p_e$ remains the same, as each edge yields an incoming (or outgoing) edge, and there are at most $N(N-1)$ incoming (or outgoing) edges, thus $p_e=\frac{z}{N(N-1)}$, as with the total degree.

In other words, let $\gamma$ and $\phi$ be integers in the intervals $1 \leq \gamma \leq 6$, $1 \leq \phi \leq 3$, and each of the basic six Erd\"os sectioning possibilities $\{E_{\gamma}\}$ have three Erd\"os sectors $E_{\gamma}= \{e_{\gamma, \phi} \}$ defined as

\begin{alignat}{3}\label{eq:part}
e_{\gamma,1}&=\{\;i\;|\;\overline{k}_{\gamma,L}\geq&&\overline{k}_{\gamma,i}\} \nonumber \\
e_{\gamma,2}&=\{\;i\;|\;\overline{k}_{\gamma,L}<\;&&\overline{k}_{\gamma,i}\leq\overline{k}_{\gamma,R}\} \\ 
e_{\gamma,3}&=\{\;i\;|\;&&\overline{k}_{\gamma,i}>\overline{k}_{\gamma,R}\} \nonumber,
\end{alignat}

\noindent where $\{\overline{k}_{\gamma,i}\}$ is

\begin{equation}
\begin{split}
\overline{k}_{1,i}&=k_i \\
\overline{k}_{2,i}&=k_i^{in} \\
\overline{k}_{3,i}&=k_i^{out} \\
\overline{k}_{4,i}&=\frac{s_i}{\overline{w}} \\
\overline{k}_{5,i}&=\frac{s_i^{in}}{\overline{w}} \\
\overline{k}_{6,i}&=\frac{s_i^{out}}{\overline{w}}
\end{split}
\end{equation}

\noindent and both $\overline{k}_{\gamma,L}$ and $\overline{k}_{\gamma,R}$ are found using $P(\overline{k})$ or $\hat{P}(\overline{k})$ as described above and illustrated in Figure~\ref{fig:setores}.

Since different metrics can be used to identify the three types of vertices, more than one metric can be used simultaneously, which is convenient when analysing small networks,
such as the cases where only 50 messages are considered in Section~\ref*{si:frac} of the Supporting Information.
%For example, a very stringent criterion can be used, according to which a vertex is only regarded as pertaining to a sector if it is so for all the metrics.
After a careful consideration of possible combinations, these were reduced to six:

\begin{itemize}
\item Exclusivist criterion $C_1$:  vertices are only classified if the class is the same according to all metrics. In this case, vertices classified do not usually reach $N$ (or 100\%), which is indicated by a black line in Figure~\ref{fig:sectIL}.

\item Inclusivist criterion $C_2$: a vertex has the class given by any of the metrics. Therefore, a vertex may belong to more than one class, and the total number of memberships may exceed $N$ (or 100\%), which is indicated by a black line in Figure~\ref{fig:sectIL}.

\item Exclusivist cascade $C_3$: vertices are only classified as hubs if they are hubs according to all metrics. Intermediary are the vertices classified either as intermediary or hubs with respect to all metrics. The remaining vertices are regarded as peripheral.

\item Inclusivist cascade $C_4$: vertices are hubs if they are classified as such according to any of the metrics. The remaining vertices are intermediary if they belong to this category for any of the metrics. Peripheral vertices are those which are classified as such with respect to all metrics.

\item Exclusivist externals $C_5$: vertices are hubs if they are classified as such according to all the metrics. Vertices are peripheral if they are peripheral or hubs for all metrics. The remaining nodes are intermediary.

\item Inclusivist externals $C_6$: hubs are vertices classified as hubs according to any metric. The remaining vertices are peripheral if they are classified as such according to any metric. The rest of the vertices are intermediary.
\end{itemize}

Using Equations~(\ref{eq:part}), these \emph{compound criteria} $C_\delta$, with $\delta$ integer in the interval $1\leq\delta\leq6$, can be specified as:

%\begin{alignat}{3}
\begin{equation}
\begin{split}
%\begin{multline}
C_1&=\{c_{1,\phi}=\left\{i\mid i\;\in e_{\gamma,\phi}, \;\forall\; \gamma\}\right\} \\
C_2&=\{c_{2,\phi}=\left\{i\mid \exists \;\;\gamma: i \in e_{\gamma,\phi}\}\right\} \\
C_3&=\{c_{3,\phi}=\left\{i\mid i\;\in e_{\gamma,\phi'}, \;\forall\; \gamma,\;\forall\;\phi'\geq \phi\}\right\} \\
C_4&=\{c_{4,\phi}=\left\{i\mid i\;\in e_{\gamma,\phi'}, \;\forall\; \gamma,\;\forall\;\phi'\leq \phi\}\right\} \\
C_5&=\{c_{5,\phi}=\left\{i\mid i\;\in e_{\gamma,\phi'}, \;\forall\; \gamma,\right.\\
&\;\;\;\;\;\;\;\;\;\;\;\;\;\;\;\;\;\; \left.\;\forall\;(\phi'+1)\%4\leq (\phi+1)\%4\}\right\} \\
C_6&=\{c_{6,\phi}=\left\{i\mid i\;\in e_{\gamma,\phi'}, \;\forall\; \gamma,\right.\\
&\;\;\;\;\;\;\;\;\;\;\;\;\;\;\;\;\;\; \left.\;\forall\;(\phi'+1)\%4\geq (\phi+1)\%4\}\right\} \\
%\end{multline}
\end{split}
\end{equation}
%\end{alignat}

Notice that the exclusivist cascade is the same sectioning of an inclusivist cascade from periphery to hubs, but with inverted order of sectors. 
The simplification of all possible compound possibilities to the small set listed above might be formalized in strict mathematical terms, but this was considered out of the scope for current interests.

%\subsubsection{Sectioning of networks in peripheral, intermediary and hubs sectors}\label{sectioning}
\subsection{Principal Component Analysis of topological metrics}\label{sec:pca}
Principal Component Analysis (PCA) is a well documented technique~\cite{pca}, used here to address the following questions:	1) which metrics contribute to each principal component and in what proportion;	2) how much of the dispersion is concentrated in each component;	3) which are the expected values and dispersions for these quantities over various networks.	This enables one to characterize human interaction networks in terms of the relative importance of network metrics and the way they combine.

Let $\mathbf{X}=\{X[i,j]\}$ be a matrix where each element is the value
of the metric $j$ at vertex $i$ .
Let
$\mu_X [j]=\frac{\sum_i X[i,j]}{I}$ be the mean of metric $j$ over all $I$ vertices, 
$\sigma_X [j]=\sqrt{\frac{\sum_i (X[i,j]-\mu_X [j])^2}{I}}$ the standard deviation of metric $j$,
and $\mathbf{X'}=\{X'[i,j]\}=\left\{\frac{X[i,j]-\mu_X[j]}{\sigma_X[j]}\right\}$ 
the matrix with the \emph{z-score} of each metric. 
Let $\mathbf{V}=\{V[j,k]\}$ be the matrix $J\times J$ of eigenvectors
of the covariance matrix $\mathbf{C}$
of $\mathbf{X'}$, one eigenvector per column.
Each eigenvector combines the original metrics into one principal component, therefore
$V'[j,k]=100\frac{|V[j,k]|}{\sum_{j'} |V[j',k]|}$
is the percentage of the principal component $k$
that is proportional to the metric $j$.
%With $k$ eigenvectors 
%$D[k]$,
%it is enough to 
Let $\mathbf{D}=\{D[k]\}$ be the eigenvalues associated with the eigenvectors $\mathbf{V}$,
then $D'[k]=100\frac{D[k]}{\sum_{k'}D[k']}$
is the percentage of total dispersion of the system that the principal component $k$
is responsible for.
We consider, in general, the three largest eigenvalues and
the respective eigenvectors in percentages:
$\{(D'[k],\;V'[j,k])\}$.
These usually sum up between 60 and 95\% of the dispersion
and reveal patterns for a first analysis.
In particular, 
given $L$ snapshots $l$ of the interaction network,
we are interested in the mean
$\mu_{V'}[j,k]$
and the standard deviation $\sigma_{V'}[j,k]$ 
of the contribution of metric $j$ to the principal component $k$,
and the mean
$\mu_{D'}[k]$
and the standard deviation 
$\sigma_{D'}[k]$
of the contribution of the component $k$ to the dispersion
of the system:

\begin{align}\label{eq:pca}
\mu_{V'}[j,k]   &=\frac{\sum_{l=1}^L V'[j,k,l]}{L}\nonumber\\
\sigma_{V'}[j,k]&=\sqrt{\frac{\sum_{l=1}^L (\mu_{V'}-V'[j,k,l])^2}{L}}\\\nonumber
\mu_{D'}[k]&=\frac{\sum_{l=1}^L D'[k,l]}{L}\\\nonumber
\sigma_{D'}[k]&=\sqrt{\frac{\sum_{l=1}^L (\mu_{D'}-D'[k,l])^2}{L}}
\end{align}

The covariance matrix 
$\mathbf{C}$ is the correlation matrix because $\mathbf{X'}$ is normalized.
Therefore, $\mathbf{C}$ is also directly observed as a first clue for patterns
by the most simple associations:
low absolute values indicate low correlation (and a possible independence);
high values indicate positive correlation;
negative values with a high absolute value indicate negative correlation.
Notice that in this case the variable $k$ is not the degree value
but a principal component.
In the results the principal components are numbered
according to the magnitude of associated eigenvalue and $k$ is incorporated into
the notation (e.g. PC2 for metrics of $\mu_{V'}[j,2]$).

\subsection{Evolution and audiovisualization of the networks}\label{sec:viz}
The evolution of the networks was observed within 
sequences of snapshots. In each sequence, a fixed number of messages,
i.e. the window size $ws$, was used for all snapshots.
%was considered with different shifts in the message timeline to obtain snapshots.
The snapshots were made disjoint in the message timeline, and were used to perform both PCA with topological metrics and Erd\"os sectioning.  
Figures and tables were usually inspected with 
$ws=\{50, 100, 200, 400, 500, 800,$ $1000, 2000, 2500, 5000, 10000\}$ messages. Variations in the number of vertices, edges
and other network characteristics, within the same window size $ws$,
are given in Section~\ref*{si:frac} of the Supporting Information document. 

Network structures were mapped to video animations, sound and musical structures developed for this research~\cite{animacoes}.% ,galGmane,appGmane}.
Such \emph{audiovisualizations} were crucial in the initial steps and
to guide the research into the most important features of network evolution.
%Furthermore, the size of the three Erd\"os sectors could be visualized in a timeline fashion.
%Visualization of network structure was especially useful in the initial inspection of data and of the structures derived from the email lists.

%This is a way to enhance the reliability of the methods,
%of algorithmic routines, of data consistency,
%and of the results themselves.
%Also, we believe that this practice raises the scientific contribution of the work,
%handing not only the framework and the results, 
%but the exact data and processes that render them~\cite{openSci}.
\section{Results and discussion}\label{sec:results}
%Remarkable features from the analysis of the four email lists are:
%\begin{itemize}
%    \item The activity along time is practically the same for all lists, thus suggesting stable patterns.
%    \item The fraction of participants in each Erd\"os sector is stable along time and can be determined even with very few messages 
%    \item The topological metrics combine into principal components in PCA in the same way for all lists and all snapshots (). 
%        \item Symmetry measures of the topology, as defined in this article, present more dispersion than the usual clustering coefficient (Section~\ref{prevalence}).
%    \item Typology speculations are immediate from results (Section~\ref{sec:pty}).
%\end{itemize}
\subsection{Activity along time}\label{constDisc}
Regular patterns of activity were observed along time
in the scales of seconds, minutes, hours, days and months.
Histograms in each of the time scales were computed as were circular average and dispersion values, and the results are given in Tables~\ref{tab:circ}-\ref{tab:min2}. For example, uniform activity is found with respect to seconds, minutes and days of the months. Weekend days exhibit about half the activity of regular weekdays, and there is a peak of activity between 11am and noon.

\begin{table}
\begin{center}
\begin{tabular}{ l|| c|c }
\hline
%scale & $\theta_\mu'$ & $S(z)$ & $Var(z)$ & $\delta(z)$ & $\frac{max(incidence)}{min(incidence)}$ & $ \mu_{\frac{max(incidence')}{min(incidence')}} $ & $ \sigma_{\frac{max(incidence')}{min(incidence')} } $ \\ \hline\hline
%& $\theta_\mu'$ & $S(z)$ & $Var(z)$ & $\delta(z)$  \\ \hline\hline
scale & mean $\theta_\mu'$ & dispersion $\delta(z)$  \\ \hline
seconds    & --//--  & 9070.17     \\
minutes    & --//--  & 205489.40   \\
hours      & -9.61   & 4.36        \\
weekdays   & -0.03   & 29.28       \\
month days & -2.65   & 2657.77     \\
months     & -0.56   & 44.00       \\\hline

\end{tabular}
\end{center}
	\caption{{\bf Time-related circular statistics.} The rescaled circular mean $\theta_\mu'$ and the circular dispersion $\delta(z)$, described in Section~\ref{sec:mtime}, for different timescales. This example table was constructed using all LAD messages, and the results are the same for other lists, as shown in Section~\ref*{si:circ} of the Supporting Information document. The most uniform distribution of activity was found in seconds and minutes. 	Hours of the day exhibited the most concentrated activity (lowest $\delta(z)$), with mean between 2 p.m. and 3 p.m. ($\theta'=-9.61$). Weekdays, days of the month and months have mean near zero (i.e. near the beginning of the week, month and year) and high dispersion. Note that $\theta_u'$ has the dimensional unit of the corresponding time period while $\delta(z)$ is dimensionless.}
\label{tab:circ}
\end{table}

\begin{table}
\footnotesize
\begin{center} 
 \begin{tabular}{ l || c | c | c | c | c | c }\hline 
  & 1h & 2h & 3h & 4h & 6h & 12h \\\hline 
 0h  &  \multirow{1}{*}{ 3.66 }   &  \multirow{2}{*}{ 6.42 }   &  \multirow{3}{*}{ 8.20 }   &  \multirow{4}{*}{ 9.30 }   &  \multirow{6}{*}{ 10.67 }   &  \multirow{12}{*}{ 33.76 }  \\\cline{2-2} 
 1h  &  \multirow{1}{*}{ 2.76 }   &   &   &   &   &  \\\cline{2-2}\cline{3-3} 
 2h  &  \multirow{1}{*}{ 1.79 }   &  \multirow{2}{*}{ 2.88 }   &   &   &   &  \\\cline{2-2}\cline{4-4} 
 3h  &  \multirow{1}{*}{ 1.10 }   &   &  \multirow{3}{*}{ 2.47 }   &   &   &  \\\cline{2-2}\cline{3-3}\cline{5-5} 
 4h  &  \multirow{1}{*}{ 0.68 }   &  \multirow{2}{*}{ 1.37 }   &   &  \multirow{4}{*}{ 3.44 }   &   &  \\\cline{2-2} 
 5h  &  \multirow{1}{*}{ 0.69 }   &   &   &   &   &  \\\cline{2-2}\cline{3-3}\cline{4-4}\cline{6-6} 
 6h  &  \multirow{1}{*}{ 0.83 }   &  \multirow{2}{*}{ 2.07 }   &  \multirow{3}{*}{ 4.35 }   &   &  \multirow{6}{*}{ 23.09 }   &  \\\cline{2-2} 
 7h  &  \multirow{1}{*}{ 1.24 }   &   &   &   &   &  \\\cline{2-2}\cline{3-3}\cline{5-5} 
 8h  &  \multirow{1}{*}{ 2.28 }   &  \multirow{2}{*}{ 6.80 }   &   &  \multirow{4}{*}{ 21.03 }   &   &  \\\cline{2-2}\cline{4-4} 
 9h  &  \multirow{1}{*}{ 4.52 }   &   &  \multirow{3}{*}{ 18.75 }   &   &   &  \\\cline{2-2}\cline{3-3} 
 10h  &  \multirow{1}{*}{ 6.62 }   &  \multirow{2}{*}{ \textbf{ 14.23 } }   &   &   &   &  \\\cline{2-2} 
 11h  &  \multirow{1}{*}{ \textbf{ 7.61 } }   &   &   &   &   &  \\\cline{2-2}\cline{3-3}\cline{4-4}\cline{5-5}\cline{6-6}\cline{7-7} 
 12h  &  \multirow{1}{*}{ 6.44 }   &  \multirow{2}{*}{ 12.48 }   &  \multirow{3}{*}{ \textbf{ 18.95 } }   &  \multirow{4}{*}{ \textbf{ 25.05 } }   &  \multirow{6}{*}{ \textbf{ 37.63 } }   &  \multirow{12}{*}{ \textbf{ 66.24 } }  \\\cline{2-2} 
 13h  &  \multirow{1}{*}{ 6.04 }   &   &   &   &   &  \\\cline{2-2}\cline{3-3} 
 14h  &  \multirow{1}{*}{ 6.47 }   &  \multirow{2}{*}{ 12.57 }   &   &   &   &  \\\cline{2-2}\cline{4-4} 
 15h  &  \multirow{1}{*}{ 6.10 }   &   &  \multirow{3}{*}{ 18.68 }   &   &   &  \\\cline{2-2}\cline{3-3}\cline{5-5} 
 16h  &  \multirow{1}{*}{ 6.22 }   &  \multirow{2}{*}{ 12.58 }   &   &  \multirow{4}{*}{ 23.60 }   &   &  \\\cline{2-2} 
 17h  &  \multirow{1}{*}{ 6.36 }   &   &   &   &   &  \\\cline{2-2}\cline{3-3}\cline{4-4}\cline{6-6} 
 18h  &  \multirow{1}{*}{ 6.01 }   &  \multirow{2}{*}{ 11.02 }   &  \multirow{3}{*}{ 15.88 }   &   &  \multirow{6}{*}{ 28.61 }   &  \\\cline{2-2} 
 19h  &  \multirow{1}{*}{ 5.02 }   &   &   &   &   &  \\\cline{2-2}\cline{3-3}\cline{5-5} 
 20h  &  \multirow{1}{*}{ 4.85 }   &  \multirow{2}{*}{ 9.23 }   &   &  \multirow{4}{*}{ 17.59 }   &   &  \\\cline{2-2}\cline{4-4} 
 21h  &  \multirow{1}{*}{ 4.38 }   &   &  \multirow{3}{*}{ 12.73 }   &   &   &  \\\cline{2-2}\cline{3-3} 
 22h  &  \multirow{1}{*}{ 4.06 }   &  \multirow{2}{*}{ 8.36 }   &   &   &   &  \\\cline{2-2} 
 23h & \multirow{1}{*}{ 4.30 }  & & & & & \\\cline{2-2}\cline{3-3}\cline{4-4}\cline{5-5}\cline{6-6}\cline{7-7} 
 \hline\end{tabular} 
 \end{center}

\label{tab:hin}
	\caption{{\bf Activity percentages along the hours of the day.} Nearly identical distributions were observed on other social systems as shown in Section~\ref*{si:hours} of the Supporting Information document.
Highest activity was observed between noon and 6pm (with 1/3 of total day activity), followed by the time period between 6pm and midnight.
Around 2/3 of the activity takes place from noon to midnight
but the activity peak occurs between 11 a.m. and 12 p.m.
This table shows results for the activity in CPP.}
\end{table}

\begin{table}
\begin{center}
\begin{tabular}{ l ||  c | c | c | c | c |   c | c}
\hline
& Mon & Tue & Wed & Thu & Fri & Sat & Sun  \\ \hline
LAU & 15.71  & 15.81  & 15.88  & 16.43  & 15.14  & {\bf 10.13}  & {\bf 10.91} \\
LAD & 14.92  & 17.75  & 17.01  & 15.41  & 14.21  & {\bf 10.40}  & {\bf 10.31} \\
MET & 17.53  & 17.54  & 16.43  & 17.06  & 17.46  & {\bf 7.92 }  & {\bf 6.06 }\\
CPP & 17.06  & 17.43  & 17.61  & 17.13  & 16.30  & {\bf 6.81 }  & {\bf 7.67 }\\\hline

\end{tabular}
\end{center}
	\caption{{\bf Activity percentages along weekdays.}
Higher activity was observed during workweek days, with a decrease of activity on weekend days of at least one third and at most two thirds.}
\label{tab:win}
\end{table}

In the scales of seconds and minutes, activity is uniform,
with the messages being slightly more evenly distributed in all lists than in simulations with the uniform distribution\footnote{Numpy version 1.8.2, ``random.randint'' function, was used for simulations, algorithms in \url{https://github.com/ttm/percolation}.}.
In the networks, $\frac{min(incidence)}{max(incidence)} \in (0.784,.794)$ while simulations reach these values but have on average more discrepant higher and lower peaks, i.e. if $\xi=\frac{min(incidence')}{max(incidence')}$ than $\mu_\xi=0.7741 \text{ and } \sigma_\xi=0.02619$.
Therefore, the incidence of messages at each second of a minute and at each minute of an hour was considered uniform.
In these cases, the circular dispersion is maximized and the mean has little meaning as indicated in Table~\ref{tab:circ}.
As for the hours of the day, an abrupt peak is found between 11am and 12pm with the most active period being the afternoon, with one third of total daily activity, and two thirds of activity are allocated in the second 12h of each day. Days of the week revealed a decrease between one third and two thirds of activity on weekends.
Days of the month were regarded as homogeneous with an inconclusive slight tendency of the first week to be more active.
Months of the year revealed patterns matching usual work and academic calendars. The time period examined here was not sufficient for the analysis of activity along the years. These patterns are exemplified in Tables~\ref{tab:hin}-\ref{tab:min2}.

\FloatBarrier

\begin{table}
\footnotesize
\begin{center}
\begin{tabular}{ l || c | c | c | c }\hline
 & 1 day & 5 & 10 & 15 days \\\hline
1 & \multirow{1}{*}{ 3.05 }  & \multirow{5}{*}{ 18.25 }  & \multirow{10}{*}{ 35.24 }  & \multirow{15}{*}{ 50.96 }  \\\cline{2-2}
2 & \multirow{1}{*}{ 3.38 }  & & & \\\cline{2-2}
3 & \multirow{1}{*}{ 3.62 }  & & & \\\cline{2-2}
4 & \multirow{1}{*}{ 4.25 }  & & & \\\cline{2-2}
5 & \multirow{1}{*}{ 3.94 }  & & & \\\cline{2-2}\cline{3-3}
6 & \multirow{1}{*}{ 3.73 }  & \multirow{5}{*}{ 16.98 }  & & \\\cline{2-2}
7 & \multirow{1}{*}{ 3.17 }  & & & \\\cline{2-2}
8 & \multirow{1}{*}{ 3.26 }  & & & \\\cline{2-2}
9 & \multirow{1}{*}{ 3.56 }  & & & \\\cline{2-2}
10 & \multirow{1}{*}{ 3.26 }  & & & \\\cline{2-2}\cline{3-3}\cline{4-4}
11 & \multirow{1}{*}{ 3.81 }  & \multirow{5}{*}{ 15.73 }  & \multirow{10}{*}{ 31.98 }  & \\\cline{2-2}
12 & \multirow{1}{*}{ 2.91 }  & & & \\\cline{2-2}
13 & \multirow{1}{*}{ 3.30 }  & & & \\\cline{2-2}
14 & \multirow{1}{*}{ 2.75 }  & & & \\\cline{2-2}
15 & \multirow{1}{*}{ 2.95 }  & & & \\\cline{2-2}\cline{3-3}\cline{5-5}
16 & \multirow{1}{*}{ 3.36 }  & \multirow{5}{*}{ 16.25 }  & & \multirow{15}{*}{ 49.04 }  \\\cline{2-2}
17 & \multirow{1}{*}{ 3.16 }  & & & \\\cline{2-2}
18 & \multirow{1}{*}{ 3.44 }  & & & \\\cline{2-2}
19 & \multirow{1}{*}{ 3.36 }  & & & \\\cline{2-2}
20 & \multirow{1}{*}{ 2.93 }  & & & \\\cline{2-2}\cline{3-3}\cline{4-4}
21 & \multirow{1}{*}{ 3.20 }  & \multirow{5}{*}{ 15.79 }  & \multirow{10}{*}{ 32.78 }  & \\\cline{2-2}
22 & \multirow{1}{*}{ 3.11 }  & & & \\\cline{2-2}
23 & \multirow{1}{*}{ 3.60 }  & & & \\\cline{2-2}
24 & \multirow{1}{*}{ 2.74 }  & & & \\\cline{2-2}
25 & \multirow{1}{*}{ 3.13 }  & & & \\\cline{2-2}\cline{3-3}
26 & \multirow{1}{*}{ 3.13 }  & \multirow{5}{*}{ 16.99 }  & & \\\cline{2-2}
27 & \multirow{1}{*}{ 3.07 }  & & & \\\cline{2-2}
28 & \multirow{1}{*}{ 3.61 }  & & & \\\cline{2-2}
29 & \multirow{1}{*}{ 3.60 }  & & & \\\cline{2-2}
30 & \multirow{1}{*}{ 3.57 }  & & & \\\cline{2-2}\cline{3-3}\cline{4-4}\cline{5-5}
\hline\end{tabular}
\end{center}

\label{tab:min}
	\caption{{\bf Activity along the days of the month.}
Nearly identical distributions are found in all systems
as indicated in Section~\ref*{si:monthdays} of the Supporting Information. Although slightly higher activity rates are found in the beginning of the month, the most important feature seems to be the homogeneity made explicit by the high circular dispersion in Table~\ref{tab:circ}.
This specific example and empirical table correspond to the activity of the MET email list.}
\end{table}

\begin{table}
\footnotesize
\begin{center}
\begin{tabular}{l || c | c | c | c | c }\hline
 & m. & b. & t. & q. & s. \\\hline
Jan & \multirow{1}{*}{ 10.22 }  & \multirow{2}{*}{ 19.56 }  & \multirow{3}{*}{ 28.24 }  & \multirow{4}{*}{ 35.09 }  & \multirow{6}{*}{ 49.16 }  \\\cline{2-2}
Fev & \multirow{1}{*}{ 9.34 }  & & & & \\\cline{2-2}\cline{3-3}
Mar & \multirow{1}{*}{ 8.67 }  & \multirow{2}{*}{ 15.53 }  & & & \\\cline{2-2}\cline{4-4}
Apr & \multirow{1}{*}{ 6.86 }  & & \multirow{3}{*}{ 20.93 }  & & \\\cline{2-2}\cline{3-3}\cline{5-5}
Mai & \multirow{1}{*}{ 7.28 }  & \multirow{2}{*}{ 14.07 }  & & \multirow{4}{*}{ 30.36 }  & \\\cline{2-2}
Jun & \multirow{1}{*}{ 6.80 }  & & & & \\\cline{2-2}\cline{3-3}\cline{4-4}\cline{6-6}
Jul & \multirow{1}{*}{ 8.97 }  & \multirow{2}{*}{ 16.29 }  & \multirow{3}{*}{ 24.47 }  & & \multirow{6}{*}{ 50.84 }  \\\cline{2-2}
Ago & \multirow{1}{*}{ 7.32 }  & & & & \\\cline{2-2}\cline{3-3}\cline{5-5}
Set & \multirow{1}{*}{ 8.18 }  & \multirow{2}{*}{ 16.25 }  & & \multirow{4}{*}{ 34.55 }  & \\\cline{2-2}\cline{4-4}
Out & \multirow{1}{*}{ 8.06 }  & & \multirow{3}{*}{ 26.36 }  & & \\\cline{2-2}\cline{3-3}
Nov & \multirow{1}{*}{ 7.64 }  & \multirow{2}{*}{ 18.30 }  & & & \\\cline{2-2}
Dez & \multirow{1}{*}{ 10.66 }  & & & & \\\cline{2-2}\cline{3-3}\cline{4-4}\cline{5-5}\cline{6-6}
\hline\end{tabular}
\end{center}

\label{tab:min2}
	\caption{{\bf Activity percentages on months along the year.} 	Activity is usually concentrated in Jun-Aug and/or in Dec-Mar, potentially due to academic calendars, vacations and end-of-year holidays. This table corresponds to activity in LAU. Similar results are shown for other lists in Section~\ref*{si:months} of the Supporting Information document.}
\end{table}

\subsection{Stable sizes of Erd\"os sectors}\label{subsec:pih}

\begin{figure*} 
\centering
\includegraphics[width=\textwidth]{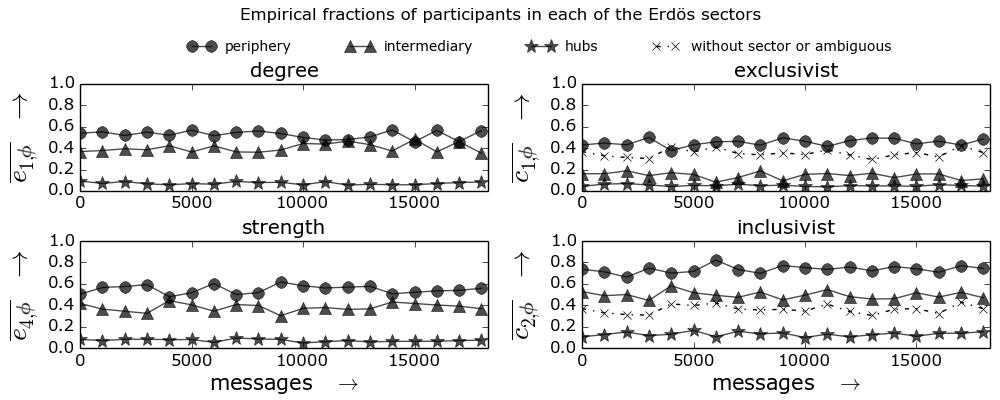}
	\caption{{\bf Stability of Erd\"os sector sizes.}
Fractions of participants derived from degree and strength criteria, $E_1$ and $E_4$ described in Section~\ref{sectioning}, are both on the left.
Fractions derived from the exclusivist $C_1$ and the inclusivist $C_2$ compound criteria are shown in the plots to the right.
The ordinates $\overline{e_{\gamma,\phi}}=\frac{|e_{\gamma,\phi}|}{N}$ denote the fraction of participants in sector $\phi$ through criterion $E_\gamma$
and, similarly, $\overline{c_{\delta,\phi}}=\frac{|c_{\delta,\phi}|}{N}$ denotes the fraction of participants in sector $\phi$ through criterion $C_\delta$.
Sections~\ref*{si:frac} and~\ref*{si:ext} of the Supporting Information bring a systematic collection of such timeline figures with all simple and compound criteria specified in Section~\ref{sectioning}, with results for networks from Facebook, Twitter and ParticipaBR.}
\label{fig:sectIL}
\end{figure*}

The distribution of vertices in the hub, intermediary, periphery Erd\"os sectors is remarkably stable along time if the snapshots hold 200 or more messages, as it is clear in Figure~\ref{fig:sectIL} and in Section~\ref*{si:frac} of the Supporting Information document. 
%Moreover, all email lists analyzed exhibit the same distribution profile.
Activity is highly concentrated on the hubs, while a very large number of peripheral vertices contribute to only a fraction of the activity.
This is expected for a system with a scale-free profile, as confirmed with the distribution of activity among participants in Table~\ref{autores}.

Typically, $[3\%-12\%]$ of the vertices are hubs,\\
$[15\%-45\%]$ are intermediary and $[44\%-81\%]$ are peripheral,
which is consistent with other studies~\cite{secFree}.
These results hold for the total, in and out degrees and strengths.
Stable sizes are also observed for 100 or less messages if the classification 
of the three sectors is performed with one of the compound criteria established in Section~\ref{sectioning}. The networks often hold this basic structure with as few as 10-50 messages, i.e. concentration of activity and the abundance of low-activity participants take place even with very few messages, which is highlighted in Section~\ref*{si:frac} of the Supporting Information. A minimum window size for the observation of more general properties might be inferred by monitoring 
both the giant component and the degeneration of the Erd\"os sectors.

In order to support the generality of these findings,
we list the Erd\"os sector sizes of 12 networks from Facebook, Twitter and ParticipaBR in Table~\ref*{tab:secE} of the Supporting Information document. The fractions of hubs, intermediary and periphery nodes are
essentially the same as for the email list networks but with exceptions and a greater variability.

\begin{table}[h]
\begin{center}
\begin{tabular}{  l ||  c | c | c | c  }
\hline
list & hub & $ Q_1 $ & $ Q_3 $ & $D_{-1}$ \\ \hline
LAU & 2.78  & 1.19 (26.35\%)  & 13.12 (75.17\%)  & 67.32 (-10.02\%) \\
LAD & 4.00  & 1.03 (26.64\%)  & 11.91 (75.18\%)  & 71.14 (-10.03\%) \\
MET & 11.14  & 1.02 (34.07\%)  & 8.54 (75.64\%)  & 80.49 (-10.02\%) \\
CPP & 14.41  & 0.29 (33.24\%)  & 4.18 (75.46\%)  & 83.65 (-10.04\%) \\\hline

\end{tabular}
\end{center}
	\caption{{\bf Distribution of activity among participants.}
The first column shows the percentage of messages sent by the most active participant. The column for the first quartile ($Q_1$) gives the minimum percentage of participants responsible for at least 25\% of total messages with the actual percentage in parentheses. Similarly, the column for the first three quartiles $Q_3$ gives the minimum percentage of participants responsible for 75\% of total messages.
The last decile $D_{-1}$ column shows the maximum percentage of participants responsible for 10\% of messages.}
\label{autores}
\end{table}

\subsection{Stability of principal components}\label{prevalence}
%The topology was analyzed using standard, well-established metrics of centrality and clustering.
%We also introduced symmetry metrics given the evidence of their importance in social contexts~\cite{newmanEvolving}.
%The contribution of each metric to the variance is very similar for all the networks and along time.

The principal components of the participants are very stable in the topological space, i.e. in the space of principal components of network measures.
Table~\ref{tab:pcain} exemplifies the formation of principal components by providing the averages over non-overlapped activity snapshots of a network. The most important result of this application of PCA, the stability of principal components, is underpinned by the very small dispersion of the contribution of each metric to each principal component.
%The contribution of each metric to the
%principal components presents
%very small standard deviation.

\begin{table}[!h]
\footnotesize
\begin{center}
\begin{tabular}{| l || c | c | c | c | c | c |}\cline{2-7}
\multicolumn{1}{c|}{} & \multicolumn{2}{c|}{PC1}          & \multicolumn{2}{c|}{PC2} & \multicolumn{2}{c|}{PC3}  \\\cline{2-7}\multicolumn{1}{c|}{} & $\mu$            & $\sigma$ & $\mu$         & $\sigma$ & $\mu$ & $\sigma$  \\\hline
$cc$ &                     0.89  & 0.59  & 1.93  & 1.33  & {\bf 21.22}  & 2.97 \\\hline
$s$ &              {\bf 11.71}  & 0.57  & 2.97  & 0.82  & 2.45  & 0.72 \\
$s^{in}$ &         {\bf 11.68}  & 0.58  & 2.37  & 0.91  & 3.08  & 0.78 \\
$s^{out}$ &        {\bf 11.49}  & 0.61  & 3.63  & 0.79  & 1.61  & 0.88 \\
$k$ &              {\bf 11.93}  & 0.54  & 2.58  & 0.70  & 0.52  & 0.44 \\
$k^{in}$ &         {\bf 11.93}  & 0.52  & 1.19  & 0.88  & 1.41  & 0.71 \\
$k^{out}$ &        {\bf 11.57}  & 0.61  & 4.34  & 0.70  & 0.98  & 0.66 \\
$bt$ &             {\bf 11.37}  & 0.55  & 2.44  & 0.84  & 1.37  & 0.77 \\\hline
$asy$ &                    3.14  & 0.98  & {\bf 18.52}  & 1.97  & 2.46  & 1.69 \\
$\mu^{asy}$              & 3.32  & 0.99  & {\bf 18.23}  & 2.01  & 2.80  & 1.82 \\
$\sigma^{asy}$           & 4.91  & 0.59  & 2.44  & 1.47  & {\bf 26.84}  & 3.06 \\
$dis$                    & 2.94  & 0.88  & {\bf 18.50}  & 1.92  & 3.06  & 1.98 \\
$\mu^{dis}$              & 2.55  & 0.89  & {\bf 18.12}  & 1.85  & 1.57  & 1.32 \\
	$\sigma^{dis}$           & 0.57  & 0.33  & 2.74  & 1.63  & {\bf 30.61}  & 2.66 \\\hline\hline
$\lambda$                & 49.56 & 1.16  & 27.14  & 0.54  & 13.25  & 0.95 \\
\hline\end{tabular}
\end{center}

\label{tab:pcain}
	\caption{{\bf Invariance of principal components.} Loadings for the 14 metrics into the principal components for the MET list, $1000$ messages in 20 disjoint positions. The clustering coefficient (cc) appears as the first metric in the table, followed by 7 centrality metrics and 6 symmetry-related metrics. Note that the centrality measurements, including degrees, strength and betweenness centrality, are the most important contributors for the first principal component, while the second component is dominated by symmetry metrics. The clustering coefficient is only relevant for the third principal component. The three components have in average more than 85\% of the variance.
The low standard deviation $\sigma$ implies that the principal components are considerably stable.}
\end{table}

The first principal component is an average of centrality metrics:
degrees, strengths and betweenness centrality.
On one hand, the similar relevance of all centrality metrics is not surprising since they are highly correlated,
e.g. degree and strength have Spearman correlation coefficient $\in [0.95,1]$ 
and Pearson coefficient $\in [0.85,1)$ for window sizes greater than a thousand messages.
On the other hand, each of these metrics is related to a different participation characteristic,
and their equal relevance for variability,
as measured by the principal component, is noticeable.
Also, this suggests that these centrality metrics 
are equally adequate for characterizing the networks
and the participants.

\begin{figure} 
\centering
\includegraphics[width=.75\textwidth]{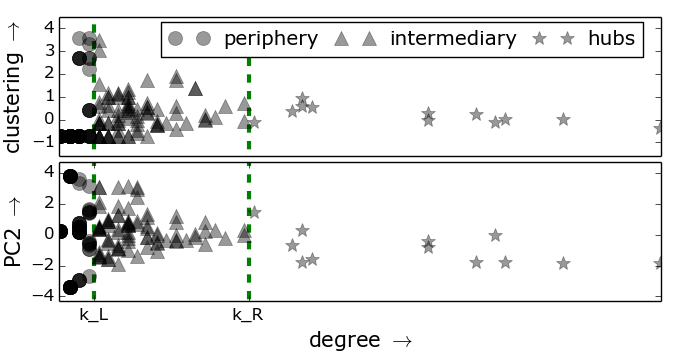}
	\caption{{\bf Symmetry-related and clustering coefficient components along connectivity.} The first plot highlights the well-known pattern of degree versus clustering coefficient, characterized by the higher clustering coefficient of lower degree vertices.
    The second plot shows the greater dispersion of the symmetry-related ordinates dominant in the second principal component (PC2).
This larger dispersion suggests that symmetry-related metrics are more powerful,
for characterizing interaction networks than the clustering coefficient,
especially for hubs and intermediary vertices.
This figure reflects a snapshot of the LAU list with 1000 contiguous messages.}

%		Similar structures were observed in all window sizes $ws\;\in\;[500,10000]$, in networks derived from email lists,
%		and in networks from Facebook, Twitter and Participabr,
%		which suggests a common relationship between the metrics of degrees, strengths and betweenness centrality,
%		the symmetry-related metrics and clustering coefficient.}
\label{fig:sym}
\end{figure}

According to Table~\ref{tab:pcain} and Figure~\ref{fig:sym},
dispersion is larger in symmetry-related metrics than in clustering coefficient.
% As expected from basic complex network theory, peripheral vertices have low values of centrality metrics and larger dispersion with regard to the clustering coefficient.
% %The scatter plot in the third system of Figure~\ref{fig:sym},
% %where all metrics are considered and there is a greater dispersion
% %with respect to the ordinates,
% This reflects in the relevance of the symmetry-related metrics.
We conclude that the symmetry metrics are more powerful, in terms of dispersion in the topological metrics space, in characterizing interaction networks and their participants, than the clustering coefficient, especially for hubs and intermediary vertices (peripheral vertices have larger dispersion with regard to the clustering coefficient).
Interestingly, the clustering coefficient is always combined
with the standard deviation of the asymmetry and disequilibrium
of edges $\sigma^{asy}$ and $\sigma^{dis}$ in the third principal component.

%These results are also reported for 12 networks from Facebook, Twitter and Participabr
%in Section~\ref{si:ext} of the Supporting Information document.
Similar results are presented in Sections~\ref*{si:pcat} and~\ref*{si:ext}
of the Supporting Information for other email lists and interaction networks. A larger variability was found for the latter networks,
which motivated the use of interaction networks derived from email lists for benchmarking.

%the overall behavior was maintained in that centrality measurements 
%were found prevalent in the first principal component,
%followed by symmetry-related metrics on the second principal
%component and then clustering coefficient on the third principal component.
%Similar results are presented in Sections~\ref{si:pcat} and~\ref{si:ext}
%of the Supporting Information document for other email lists and other interaction networks,
%with the consideration of strategic combinations of metrics.

\subsection{Types from Erd\"os sectors}\label{sec:pty}

Assigning a type to a participant raises important issues about the scientific cannon for human types and the potential for stigmatization and prejudice. The Erd\"os sector to which a participant belongs can be regarded as implying a social type for this participant.
In this case, the type of a participant changes both along time and as different networks are considered, despite the stability of the network. Therefore, the potential for prejudice of such participant typology is attenuated~\cite{adorno}. In other words, an individual is a hub in a number of networks and peripheral in other networks, and even within the same network he/she most probably changes type along time~\cite{animacoes}.

\begin{figure}
\centering
\includegraphics[width=.7\columnwidth]{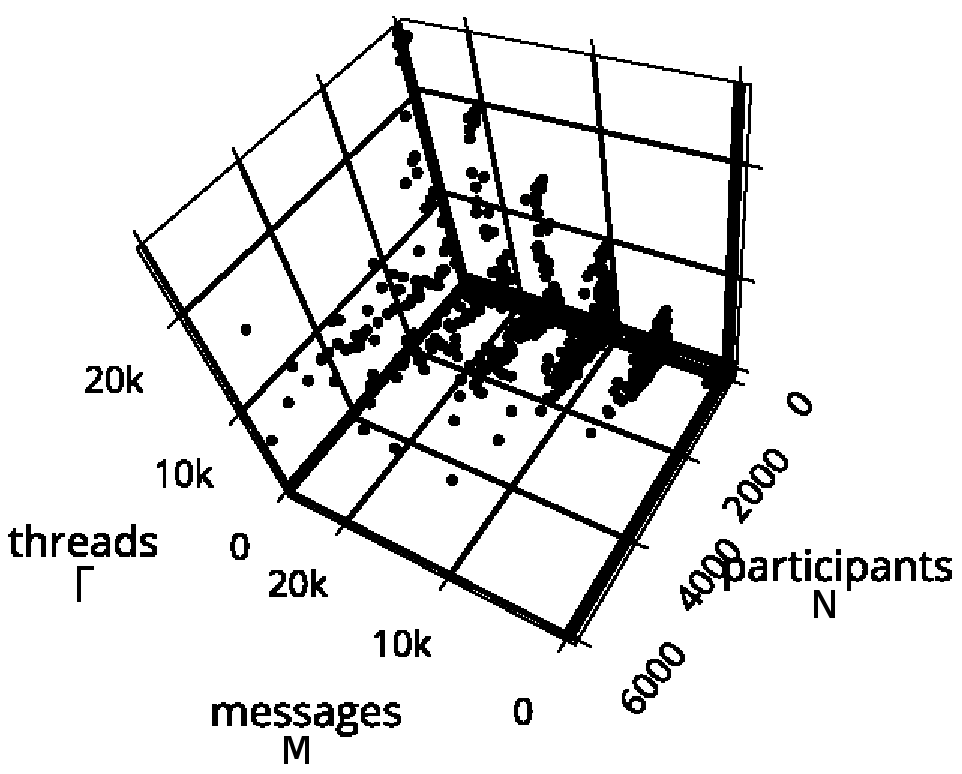}
	\caption{{\bf Threads against participants and messages.} A scatter plot of number of messages $M$ versus number of participants $N$ versus number of threads $\Gamma$ for 140 email lists.
Highest $\Gamma$ is associated with low $N$.
The correlation between $N$ and $\Gamma$ is negative for low values of $N$ but positive otherwise.
This negative correlation between $N$ and $\Gamma$ can also be observed in Table~\ref{tab:genLists}.
Accordingly, for $M=20000$ messages, this inflection
of correlation was found around $N=1500$, while CPP, LAU, LAD, MET lists 
present smaller networks.}
\label{fig:nmgamma3d}
\end{figure}

The importance of this issue can be grasped by the consideration of static types derived from quantitative criteria. For example, in email lists with a small number of participants, the number of threads has a negative correlation with the number of participants.
When the number of participants exceeds a threshold, the number of threads has a positive correlation with the number of participants.
This finding is illustrated in Figure~\ref{fig:nmgamma3d}
and can also be observed in Table~\ref{tab:genLists}.
The assignment of types to individuals, in this latter case,
has more potential for prejudice because
the derived participant type is static and
one fails to acknowledge that
human individuals are not immutable entities.

Further observations regarding the Erd\"os sectors
and the implicit participant types were made, which are consistent with the literature~\cite{barabasiEvo}: 1) hubs and intermediary participants usually have intermittent activity, and stable activity was found only in smaller communities. For instance, the MET list had stable hubs while LAU, LAD and CPP exhibited intermittent hubs.
2) Network structure seems to be most influenced by the
activity of intermediary participants as they have less extreme
roles than hubs and peripheral participants and
can therefore connect to the sectors and other participants 
in a more selective and explicit manner.

%\section{Discussion}
% given the results, and before reaching the conclusions
% what to say?
% --> what is the overall knowledge derived from the results
% --> what are the limitations of this knowledge and of individual results
% --> how should this results carry on is on the next sections.
%\subsection{Consecutive scientific research}
% --> research
% textual diferences
% audiovisualization of data
% typologies, sociological critical theory, social psychology

%\subsection{Technological applications}
% --> technological 
% resources categorization and recommendation
% document creation
% ontologies for the semantic web

%\subsection{Experimental and theoretical aspects of the research}
% --> methods
% Exploratory?
% Hypothesis testing?
% --> contributions
% verifiable
% knowledge
% contextualization in the academic knowledge

\subsection{Implications of the main findings}\label{sec:impl}
The findings reported in this article arose from an exploratory procedure to visually inspect the networks and to analyze considerable amounts of interaction networks data.
% deriving from email lists and also from other networks.
While this procedure has certainly an ad hoc nature, the statistics in the data are sufficiently robust for important features from these interaction networks to be extracted.
Temporal stability, in the sense that interaction networks could be considered as stationary time series, is the most important feature. Also relevant is the significant stability found on the principal components, on the fraction of participants in each Erd\"os Sector and on the activity along different timescales. In fact, these findings confirm our initial hypothesis - based on the literature~\cite{newmanBook} - that interaction networks should exhibit some stability traces. The potential generality of these findings is suggested by the analysis of networks derived from diverse systems, with interaction networks from public email lists serving as proper benchmarks. Indeed, with such benchmarks one can compare any social network system. Furthermore, this analysis enables us to establish an outline of human interaction networks. It takes the hub, intermediary and periphery sectors out of the scientific folklore and into classes drawn from quantitative criteria. It enables the conception of non-static human types derived from natural properties.

 
We envisage that the knowledge generated in the analysis may be exploited in applications where the type of each participant and the relative proportion of participants in each sector can be useful metadata. Just by way of illustration, this could be applied in semantic web initiatives, given that the Erd\"os sectorialization is static in a given snapshot. These results are also useful for classifying resources, e.g. in social media, and for resources recommendation to users~\cite{opa}. 
Finally, the knowledge acquired with a quantitative treatment of the whole data may help guide the creation through collective processes of documents to assist in participatory democracy.

 
Perhaps the most outreaching implications are related to sociological consequences. The results expose a classification of human individuals which is directly related to the concentration of wealth and based on natural laws. The derived human typology changes over different systems and over time in the same system, which implies a negation of the absolute concentration of wealth. Such concentration exists but changes across different wealth criteria and with time. Also, the hubs stand out as dedicated, sometimes enslaved,
components of the social system. The peripheral participants have very limited interaction with the network. This suggests that intermediary participants tend to dictate structure, legitimate the hubs and stand out as authorities.

 
With regard to the limitations of our study, one should emphasize that not all types of human interaction networks were analyzed. Therefore, the plausible generalization of properties has to be treated with caution, as a natural tendency of such systems and not as a rule. Also, the stable properties in the networks were not explored to the limit, which leaves many open questions. For example, what are the maximum and minimum sizes of the networks for which they hold? What is the outcome of PCA analysis when more metrics are considered? What is the granularity in which the activity along the timescales is preserved? Do the findings reported also apply to other systems, beyond human networks?

\section{Conclusions}\label{sec:conc}
The very small standard deviations of principal components formation
(see Sections~\ref{sec:pca} and~\ref{prevalence}),
the presence of the Erd\"os sectors even in networks with
few participants (see Sections~\ref{sectioning} and~\ref{subsec:pih}),
and the recurrent activity patterns along different timescales (see Sections~\ref{sec:mtime} and~\ref{constDisc}),
go a step further in characterizing scale-free networks in the context
of the interaction of human individuals.
Furthermore, the importance of symmetry-related metrics,
which surpassed that of clustering coefficient,
with respect to dispersion of the system in the topological measures space,
might add to the current understanding of key-differences between digraphs and
undirected graphs in complex networks.
Noteworthy is also the very stable fraction participants in each Erd\"os sector when the network reaches more than 200 participants.
Benchmarks were derived from email list networks
and the supplied analysis of
networks from Facebook,
Twitter and ParticipaBR in the Supporting Information might ease hypothesizing
about the generality of these characteristics.

Further work should expand the analysis to include
more types of networks and more metrics.
The data and software needed to attain these results
should also receive dedicated and in-depth
documentation as they enable a greater level of transparency
and work share,
which is adequate for both benchmarking
and specifically for the study of systems constituted
by human individuals (see Section~\ref{sec:data}).
The derived typology of hub, intermediary and peripheral participants
has been applied for semantic web and participatory democracy efforts,
and these developments might be enhanced to yield scientific knowledge~\cite{opa}.
Also, we plan to further explore and publish the audiovisualizations
used for this research~\cite{versinus,animacoes} and
the linguistic differences found in each of the Erd\"os sectors~\cite{rcText}.

% trabalhos de visualização (versinus), diferenciação de texto
% aplicação da tipologia para democracia participativa e dados ligados
% necessidade de publicar os dados em formatos ligados para comparacao da estrutura
% Future work on including more measures, other networks, sharing
% data in RDF for benchmarking, exploring larger timescales
% and of the data and OWL ontologies 
% 
% 

%A systematic study of the activity of participants belonging to the three 
%distinct Erd\"os sectors indicated simple patterns for hubs and peripheral vertices,
%while the network structure was governed by the intermediary vertices.
%These properties were shared by all email lists and were time-independent,
%which is consistent with the literature.
%We may therefore consider the Erd\"os sectors as leading to a human typology which bridges exact sciences, with quantitative procedures for the classification, and human sciences, where there is a legacy in the observation of human types. 

\subsection{Acknowledgments}
Financial support was obtained from CNPq (140860/2013-4,
project 870336/1997-5), United Nations Development Program (contract: 2013/000566; project BRA/12/018) and FAPESP. 
The authors are grateful to the American Jewish Committee for maintaining an online copy of the Adorno book used on the epigraph~\cite{adorno}, to Gmane creators and maintainers for the public email list data, to the communities of the email lists and other groups used in the analysis, and to the Presidency of the Brazilian Republic for keeping ParticipaBR code and data open.
We are also grateful to developers and users of Python scientific tools, to Leonardo Paulo Maia (IFSC/USP) and to Francisco J. P. Lopes (UFRJ) for valuable insights.

\section*{References}

\bibliography{paper}

\end{document}